# Analysis of Eddy Current Distributions in the CMS Magnet Yoke During the Solenoid Discharge

V. I. Klyukhin, *Member, IEEE*, D. Campi, B. Curé, A. Gaddi, H. Gerwig, J. P. Grillet, A. Hervé, R. Loveless, and R. P. Smith

*Abstract*—Flux loops have been installed on selected segments of the magnetic flux return yoke of the 4 T superconducting coil of the Compact Muon Solenoid (CMS) detector under construction at CERN. Voltages induced in the loops during discharge of the solenoid will be sampled online during the entire discharge and integrated offline to provide a measurement of the initial magnetic flux density in steel at the maximum field to an accuracy of a few percent. Although the discharge of the solenoid is rather slow (190 s time constant), the influence of eddy currents induced in the yoke elements should be estimated. The calculation of eddy currents is performed with Vector Fields' program ELEKTRA. The results of the calculations are reported.

## I. INTRODUCTION

THE compact muon solenoid (CMS) is a general-purpose detector designed to run at the highest luminosity at the CERN Large Hadron Collider (LHC). Its distinctive features include a 4 T superconducting solenoid with 6-m diameter by 12.5-m long free bore, enclosed inside a 10 000-ton yoke made of construction steel: five dodecagonal three-layered barrel wheels and three end-cap disks at each end, comprised of steel plates up to 620-mm thick, which return the flux of the solenoid and serve as the absorber plates of the muon detection system [1], [2].

A three-dimensional (3-D) model of the magnetic field of the CMS magnet has been prepared [3] for utilization during the engineering phase of the magnet system and early physics studies of the anticipated performance of the detector, as well as for track parameter reconstruction when the detector begins operation.

To reduce the uncertainty in utilization of the calculated values for the magnetic field, which is used to determine the momenta of muons during detector operation, a direct measurement of values of the average magnetic flux density in selected regions of the yoke by an integration technique is planned with 22 405-turn flux-loops installed around selected CMS yoke plates.

The areas enclosed by the flux-loops vary from 0.28 to 1.53 $m^2$ on the barrel wheels, and from 0.48 to 1.1 $m^2$ on the end-cap disks. The flux-loops will measure the variations of the magnetic flux induced in the steel when the field in the solenoid is changed during the "fast" (190 s time constant) discharge made possible by the protection system provided to protect the magnet in the event of major faults [4], [5]. The protection system will be tested during the commissioning of the CMS magnet that will provide the opportunity for the flux-loop measurements. Neither the measurements with flux-loops during the solenoid charge-up (that requires several hours) nor reversing the solenoid current are planned for the commissioning of the CMS magnet.

To investigate if the measurements of the average magnetic flux density in the CMS yoke plates could be done with accuracy of a few percent using flux-loops, a special R&D program was performed with sample disks made of the CMS yoke steel from different melts [6], [7]. This steel contains up to 0.17% C, up to 1.22% Mn, and also some Si, Cr, and Cu.

The disks were placed by turns into a slowly increasing and decreasing external magnetic flux, which was produced by a laboratory dipole electromagnet connected to a computer-controlled power supply. The measured voltages were induced in a test flux-loop mounted on the disk. These studies indicated the magnetic flux density in a steel object magnetized by an external source could be measured with good precision using a combination of the flux-loop and Hall probes mounted on the surface of the steel. The Hall probes measured the remanent field on the disk steel-air interface after discharge of the electromagnet and this field was taken into account. The studies also showed that the contribution of eddy currents to the voltages induced in the test flux-loop is negligible.

In this paper, a study is undertaken to estimate the contribution of eddy currents to the voltages induced in the flux-loops installed on the CMS magnet yoke when the "fast" discharge of the CMS coil occurs. In Section II, the CMS magnet models, used in the calculations, are described. In Section III, the results of the calculations are presented. In Section IV, the forces acting on brass absorbers of the CMS hadronic barrel (HB) and end-cap (HE) calorimeters during the "fast" discharge are reported, and the conclusions from the study are presented in Section V.

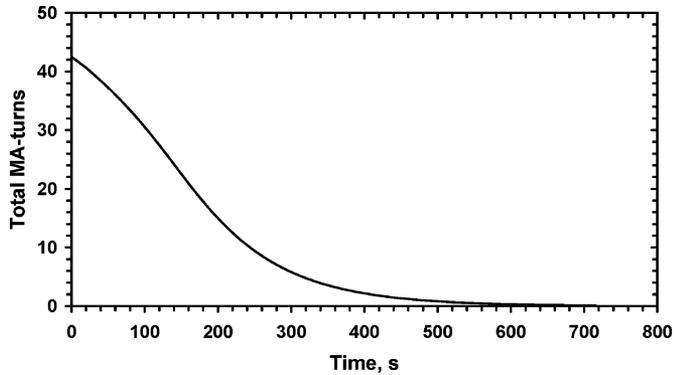

Fig. 1. Calculated CMS coil "fast" discharge.

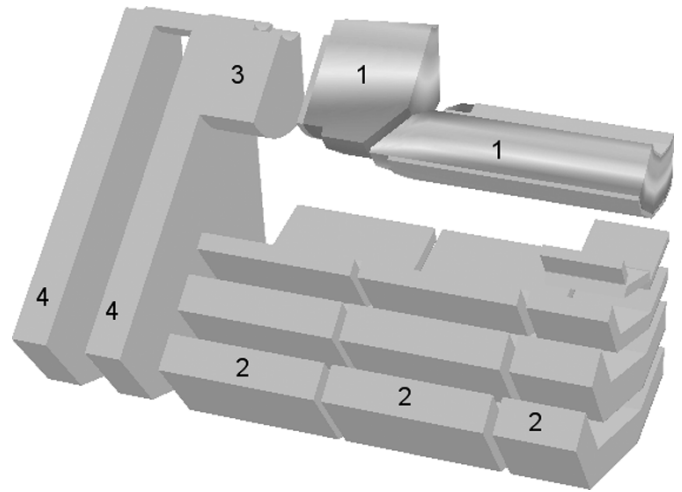

Fig. 3. ELEKTRA model used for the calculation of eddy currents in the brass absorbers of the hadronic calorimeters. (1) The segments of the brass absorbers, (2) barrel wheels, (3) nose disk, and (4) end-cap disks are displayed. The superconducting solenoid included in the model is not shown.

To perform ELEKTRA analysis of eddy currents in the CMS yoke at 15 output times 415 CPU hours on a 450 MHz processor machine was required. To meet the batch queue requirements and to vary the time step, the analysis restarted at 50, 100, 150, 200, and 300 s. To analyze the magnetic field distribution in absence of eddy currents at the same output times, 12.3 CPU hours were required. To calculate the eddy currents in the brass absorbers of the HB and HE calorimeters, 61.4 CPU hours were required.

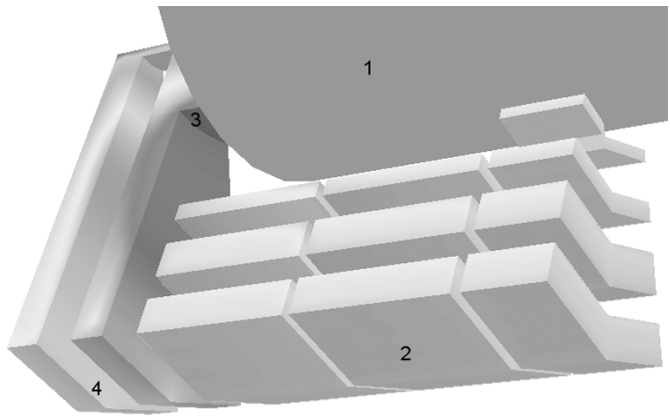

Fig. 2. ELEKTRA model used for the yoke eddy current calculation. (1) The CMS coil, (2) the segments of the barrel wheels, (3) nose disk, and (4) end-cap disks are shown.

## II. ELEKTRA MODELS USED IN THE CALCULATIONS

The "fast" discharge of the CMS solenoid will cause quenching of the superconducting coil and the current decay departs from a simple L/R(t) decay of an inductor (L) into an external resistance (R)(t) changing with time (t) as simulated and shown in Fig. 1 [5]. The derivative of the current with respect to time reaches an extreme at 140 s from a start of the discharge.

This discharge results in flux changes in various parts of the steel yoke that, in turn, causes the eddy currents in the elements of the CMS steel yoke and in brass absorbers of the CMS HB and HE calorimeters. To estimate if these parasitic currents could influence to the voltages induced in the flux-loops in different parts of the yoke, and also to calculate the forces acting on brass absorbers of the hadronic calorimeters, the 3-D models of the CMS magnet displayed in Figs. 2 and 3 are investigated with Vector Fields' program ELEKTRA [8].

Calculations with ELEKTRA, which utilizes a vector potential in the regions where the eddy currents are expected, are very CPU time consuming. To reduce CPU time to a reasonable amount, the CMS yoke is described in a simplified way, the number of finite element nodes in the models is reduced to a reasonable value, the time step varied from 6.25 to 25 s, and the number of output times in the transient analysis of the current decay following the drive function shown in Fig. 1 did not exceed 15.

### A. Models for Analysis of Eddy Currents in the CMS Yoke

The model of the CMS magnet shown in Fig. 2 is used to perform ELEKTRA analysis of eddy currents in the CMS yoke.

This model includes the entire CMS superconducting coil at cryogenic temperature and a 1/24 segment of the yoke that is then rotated and reflected in the OPERA-3d [8] postprocessor analysis to obtain the full description of the CMS yoke. This 30° azimuthal segment of the yoke is described as two and one-half three-layered barrel wheels, a small nose disk, and two thick end-cap disks. Neither the connection brackets between the barrel layers nor the azimuth gaps in the CMS barrel wheels are modeled. The thin end-cap disks and ferromagnetic parts of the CMS forward hadronic calorimeter are also omitted. This leads to a small overestimation of the eddy currents in the yoke cross-sections where the flux-loops are located.

Different magnetic and electrical properties of materials are used to describe three different regions of the yoke: the tail catcher (the short steel barrel at minimum radius at the median plane of the system) and the first full-length thin barrel layer (region 1); second and third thick barrel layers (region 2); the nose and end-cap discs (region 3).

A vector potential is used in all three regions. The electrical resistivity of construction steel used in calculations in regions 1, 2, and 3 is equal to 0.18, 0.15, and 0.165 $\mu\Omega \cdot$ m respectively.

To eliminate the eddy current influence on the voltages simulated in the flux-loops, the same geometry of the CMS magnet is used in another ELEKTRA model. This model assumes an infinite electrical resistivity and total scalar magnetic potential instead of vector potential in all regions of the yoke.

Both models give the values of the magnetic flux density in the center of the CMS coil that agrees at the same times within 0.6% in average.

### B. Model for Calculation of Forces on the Calorimeter Absorbers

The eddy currents in the brass absorbers of the HB and HE calorimeters are calculated with the model presented in Fig. 3. This model includes the entire CMS superconducting coil at cryogenic temperature and a 1/12 segment of the yoke of the same configuration as described above. The model also includes a 1/12 segment of the hadronic calorimeter brass absorbers together with inner, outer, and back support plates of the calorimeters made of stainless steel. A 60° azimuthal segment is the minimum configuration, which satisfies rotational symmetry of the different azimuth segmentations of the yoke and absorbers.

The model assumes an infinite electrical resistivity and total scalar magnetic potential in all regions of the yoke. The electrical resistivity of brass is equal to $0.06158\ \mu\Omega \cdot \mathrm{m}$ in the HB calorimeter absorber, and $0.09610\ \mu\Omega \cdot \mathrm{m}$ in the HE calorimeter absorber. The electrical resistivity of stainless steel is taken to $0.6897\ \mu\Omega \cdot \mathrm{m}$.

## III. CALCULATION OF EDDY CURRENTS IN THE CMS YOKE

The calculations of eddy currents in the CMS yoke are performed with ELEKTRA at 0, 25, 50, 100, 125, 150, 175, 200, 250, 300, 350, 400, 500, 600, and 700 s from the start of the simulated CMS "fast" discharge. At the same output times another ELEKTRA analysis is done with the model, which assumes an infinite electrical resistivity and total scalar magnetic potential instead of vector potential in all regions of the yoke.

The maximum eddy current density is investigated in the yoke cross-sections where the 22 flux-loops are located. The calculation indicates that the maximum eddy currents in the yoke barrel cross-sections arrive at 140 s after the beginning of the discharge, the same time when the derivative of the current with respect to time reaches an extreme. The eddy currents in the yoke end-cap disk cross-sections reach the maximum approximately 20 s later.

In the cross-section of the tail catcher enclosed by the flux-loop the maximum eddy current density is $2.59\ \mathrm{kA/m^2}$. In the cross-sections of the first thin barrel layer enclosed by the flux-loops the maximum eddy current density varies from 4.16 to $12.9\ \mathrm{kA/m^2}$. In similar cross-sections of second barrel layer the maximum eddy current density varies from 5.14 to $12.5\ \mathrm{kA/m^2}$. In the cross-sections of third barrel layer the maximum eddy current density varies from 5.42 to $7.38\ \mathrm{kA/m^2}$.

In the first end-cap disk cross-sections enclosed by the flux loops the maximum eddy current density varies from 27.12 to $51.98\ \mathrm{kA/m^2}$, and in second end-cap disk cross-sections the maximum eddy current density varies from 11.21 to $17.52\ \mathrm{kA/m^2}$.

Fig. 2 shows that the absolute maximum eddy current density is distributed over the surfaces of the nose disk and the first end-cap disk nearest the CMS coil, and on the surfaces surrounding the beam pipe.

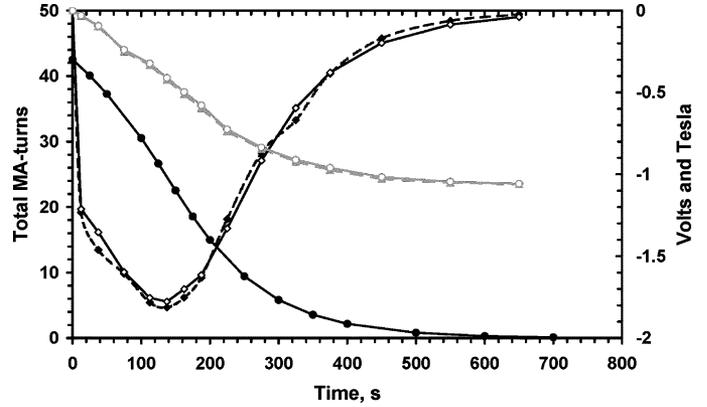

Fig. 4. Voltages calculated in the flux-loop on the thick block in the second barrel layer when the eddy currents with realistic electrical resistances (solid black line with white diamonds) and infinite resistances (smoothed dotted black line with black diamonds) are modeled during the current ramp (dark grey line with dark grey circles). The solid grey line with white circles represents the result of voltage integration when the eddy currents exist. The dotted grey line with grey triangles displays the result of voltage integration in the model with eddy currents suppressed. The difference between two integrated magnetic flux densities is 0.29%.

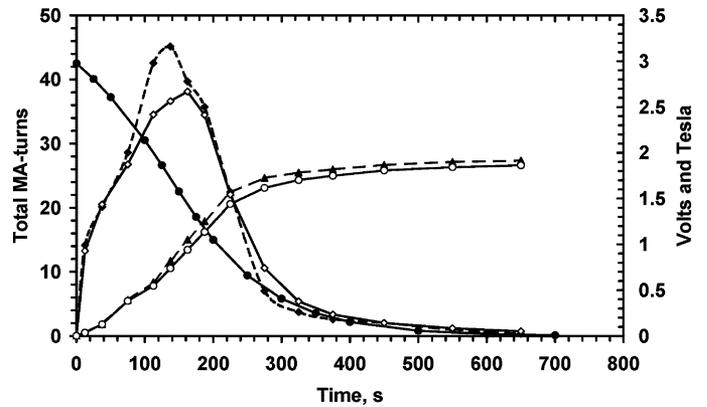

Fig. 5. Voltages calculated in the flux-loop on 18° segment of the second end-cup disk when eddy currents from realistic electrical resistances (solid black line with white diamonds) and eddy currents suppressed by infinite resistances (smoothed dotted black line with black diamonds) are modeled during the current ramp (dark grey line with dark grey circles). The solid grey line with white circles represents the result of voltage integration when eddy currents exist. The dotted grey line with grey triangles displays the result of voltage integration when eddy currents are suppressed. The difference between two integrated magnetic flux densities is 2.76%.

To investigate if these values of the eddy current density change the magnetic flux $\Phi$, and thus, the average magnetic flux density in the yoke cross-sections enclosed by the flux-loops, the average voltages $\mathrm{V} = \Delta\Phi/\Delta\mathrm{t}$ induced in the flux-loops by the magnetic flux changes between time intervals are calculated as shown in Figs. 4 and 5.

The voltages are calculated in both models—the first with realistic eddy currents from realistic electrical resistances and the second with eddy currents suppressed by infinite electrical resistances. To obtain valid comparison of the voltages, the normalization of the magnetic flux values is done at zero time. In the analysis with the realistic eddy currents the maximum amplitudes of the voltages calculated in the flux-loops on the barrel blocks range from 0.4 to 2.14 V. The voltages in the flux-loops on the segments of the end-cap disks range from 1.74 to 2.7 V.

The voltages obtained in both models are integrated by multiplying the average voltage in each time interval by the length of time interval. The time integrals of the voltages are the total flux changes in the flux-loops. The obtained flux values are renormalized to magnetic flux density using the areas of the flux-loops and the numbers of turns in the flux loops.

An estimation of the eddy current influence on the calculated values of the average magnetic flux density is obtained by comparison of those values integrated in the models with and without eddy currents.

The expected average eddy current contributions are as follows: $(0.22 \pm 0.89)\%$ in the yoke cross-sections enclosed by the flux-loops on the barrel wheels; $(-0.83 \pm 2.42)\%$ in the yoke cross-sections enclosed by the flux-loops on the end-cap disks; and $(-0.067 \pm 1.55)\%$ in all the yoke cross-sections enclosed by the flux-loops. A minus sign indicates that the value of the average magnetic flux density integrated in the model with eddy currents is less than the same value in the model without eddy currents.

These contributions lie well within the expected uncertainties of 2–3% anticipated in the flux-coil measurements of the average magnetic flux density in steel elements of the CMS yoke, as determined in the R&D program [6], [7].

## IV. Calculation of Forces on the Brass Absorbers of the CMS Hadronic Calorimeters

The calculations of eddy currents in the brass absorbers of the HB and HE calorimeters are performed with ELEKTRA at 0, 25, 50, 100, 125, 150, 175, and 200 s from the start of the simulated CMS "fast" discharge.

The distribution of the calculated eddy current density is shown in Fig. 3. The maximum eddy current density of 135 $kA/m^2$ is reached in the brass absorber of the HB calorimeter between 125 and 150 s after beginning of the "fast" discharge. The maximum eddy current density of 157 $kA/m^2$ occurs in brass absorber of the HE calorimeter at the same time.

The volume integration of the components of the vector product of eddy current density and magnetic flux density vectors gives the axial and radial components of the forces acting on the segments of the calorimeter absorbers.

The maximum axial force of $-26.4$ kN on a half segment of the HB calorimeter absorber subtending 60° in azimuth is reached 100 s after the beginning of the fast discharge and is directed to the CMS coil center. At that time the radial force directed to the outer radius of the HB calorimeter absorber is 787.5 kN on a segment subtending 60° in azimuth.

The maximum axial force of $-59.0$ kN on a half segment of the HE calorimeter absorber subtending 60° in azimuth is reached 50 s after the beginning of the fast discharge and is directed to the CMS coil center. At that time the radial force directed to the outer radius of the HE calorimeter absorber is 824.9 kN on a segment subtended 60° in azimuth.

All the computed forces are well within the safety margins of the calorimeter mounting systems.

## V. Conclusion

The estimated influence of the eddy currents on measurements of the average magnetic flux density in ferromagnetic parts of the CMS yoke that will be performed during commissioning the CMS magnet is at the level of a few percent, consistent with the demonstrated accuracy of the flux-loop measurement technique.

The computed forces on the brass absorbers of the CMS hadronic calorimeters originated by the eddy currents during the CMS "fast" discharge are within the safety margins of the mounting systems.

The analysis performed in this study confirms the possibility to use the "fast" discharge of the CMS coil for measurements of the average magnetic flux density in ferromagnetic parts of the CMS yoke without undue influence from eddy currents.

## References

[1] A. Hervé, G. Acquistapace, D. Campi, P. Cannarsa, and P. Fabbricatore et al., "Status of the CMS magnet," *IEEE Trans. Appl. Supercond.*, vol. 12, no. 1, pp. 385–390, Mar. 2002.
[2] A. Hervé, B. Blau, D. Campi, P. Cannarsa, B. Curé, and T. Dupont et al., "Status of the construction of the CMS magnet," *IEEE Trans. Appl. Supercond.*, vol. 14, no. 2, pp. 542–547, Jun. 2004.
[3] V. I. Klioukhine, D. Campi, B. Curé, A. Desirelli, S. Farinon, and H. Gerwig et al., "3D magnetic analysis of the CMS magnet," *IEEE Trans. Appl. Supercond.*, vol. 10, no. 1, pp. 428–431, Mar. 2000.
[4] V. I. Klioukhine and R. P. Smith, On a Possibility to Measure the Magnetic Field Inside the CMS Yoke Elements, CERN, Geneva, Switzerland, CMS Internal Note 2000/071, Nov. 2000.
[5] B. Cure and C. Lesmond, "Synthesis on Fast Discharge Studies," CEA/Saclay, Saclay, France, DSM/DAPNIA/STCM Tech. Report 5C 2100T-1000 032 98, Nov. 1999.
[6] R. P. Smith, D. Campi, B. Curé, A. Gaddi, H. Gerwig, J. P. Grillet, A. Hervé, V. Klyukhin, and R. Loveless, "Measuring the magnetic field in the CMS steel yoke elements," *IEEE Trans. Appl. Supercond.*, vol. 14, no. 2, pp. 1830–1833, Jun. 2004.
[7] V. I. Klyukhin, D. Campi, B. Curé, A. Gaddi, H. Gerwig, J. P. Grillet, A. Hervé, R. Loveless, and R. P. Smith, "Developing the technique of measurements of magnetic field in the CMS steel yoke elements with flux-loops and hall probes," *IEEE Trans. Nucl. Sci.*, vol. 51, no. 5, pp. 2187–2192, Oct. 2004.
[8] *ELEKTRA/OPERA-3d Software*, Vector Fields Ltd., Oxford, U.K.